# Screening in coupled low-dimensional systems: an effective polarizability picture

Ting-Ting Kang

*National Laboratory for Infrared Physics, Shanghai Institute of Technical Physics, Chinese Academy of Sciences, 200083 Shanghai, People's Republic of China*

The screening of an individual low-dimensional object can be strongly influenced by the objects nearby. We propose that such environment's influence can be absorbed into an effective polarizability, instead of its intrinsic polarizability. Using a toy system consists of two spatially separated 2DEG layers gas an example, this picture is analytically deduced via a multi-component RPA theoretical method. We show that the resultant effective polarizability of backside layer is just the dielectric function describing the system's collective plasmon excitations. Furthermore, several interesting topics are discussed, e.g. the mutual modulation of Friedel oscillation, the ultimate screening limit imposed by metal layer, smoothing potential inhomogeneity in back screening configuration.

Screening, namely the damping of electric fields caused by the presence of mobile charge carriers, is of greatly scientific and technological importance [1]. Interestingly, metal and insulator represent two opposite extremes of screening: static electric field can penetrate insulator without attenuation, but be completely shielded by metal. What makes difference is that, at Fermi level, density of state (DOS=$\partial n/\partial \mu$, $n$: electron density, $\mu$: chemical potential) of insulator is zero, so no carrier is available for screening. While metal has an infinite number of carriers for screening due to its infinite large ($\infty$) DOS. Other useful concept, like compressibility $K$ ($K^{-1} = n^2 \partial\mu/\partial n$) [2, 3], can also derive from DOS.

Regarding most heavily studied low-dimensional systems, such as 2DEG (two dimensional electron gas) in AlGaAs/GaAs, graphene [4], LaAlO$_3$/SrTiO$_3$ [5], they all have limited DOS, namely not zero but also not $\infty$, leading to an remarkable phenomenon- partial field penetration. Partial field penetration is not trivial, because it depends on not only the studied object, but also the objects nearby, even those spatially separated objects are in behind! So a fundamental consequence is resulted: although sometimes only the screening of one certain object is interested, we should consider all the objects together as a coupled system. This situation looks like the environment contribute some DOS to the studied object's DOS. An attractive strategy is to absorb the influence of environment into the studied object, leading to an imagined "effective" DOS. Therefore it only needs to consider this modified "effective" object alone. However, DOS and compressibility $K$ is not sufficient if the spatial variation is included. Being a function of wave vector $q$, polarizability $\Pi(q)$ [with $\Pi(0)$ = DOS] can extend this "effective DOS" picture to more complicated geometric configuration, giving rise to an "effective polarizability" picture (EPP). And our paper is largely motivated by this idea and wants to provide some fundamental insights.

To capture the generic screening physics among low-dimension objects, the coupled system in which we are interested contains two 2DEG (two-dimension electron gas) layers, which can be semiconductor quantum wells, graphene or metal layer. A point charge above the upper layer, provide the needed non-uniform electric field. The choosing of this toy system is highly meaningful. First, point charge produces an in-plane inhomogeneous potential, which can model a disordered potential. The screening (or smoothening) of such disordered potential would find many applications, like the charge-impurity induced potential disorder in graphene [6]. Second, this system is the most accessible in experiments. For example, AlGaAs/GaAs DQW is very mature in experimental realization. And Eisenstein *et al.*[2] had used it to experimentally prove the existence of partial field penetration.

Before our research, the widely used theoretical method for screening in inhomogeneous system is Thomas-Fermi (T-F) model.[7]. T-F model is constructed on two simple equations, i.e. Poisson's equation [Eq.(1.1)] and a linear relation between chemical potential $\mu$ and local carrier density $n(\vec{R})$ [Eq.(1.2)]:

$$\nabla^2 \mu(\vec{R}) = -\frac{[n(\vec{R}) - n_0]}{\varepsilon_s} \quad (1.1)$$

$$n(\vec{R}) - n_0 = N(\mu)[\mu(\vec{R}) - \mu_F] \quad (1.2)$$

where $n_0$: the equilibrium carrier density; $N(\mu)$: DOS at energy $\mu$; $\mu_F$: the Fermi energy; $\varepsilon_s$: the dielectric constants of hosting matrix. SI units (instead of Gauss units) are used throughout this paper.

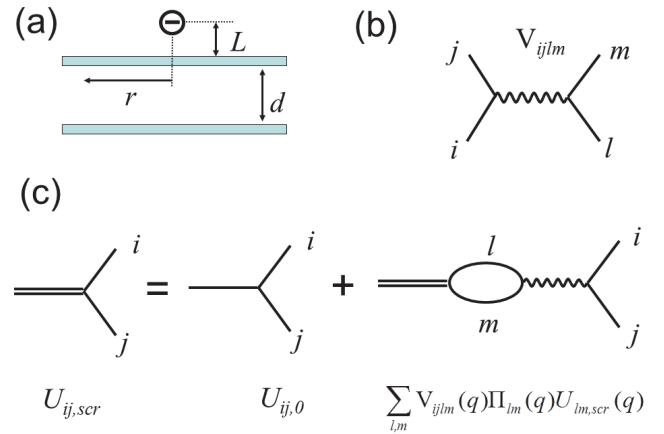

Fig. 1 (color online). (a) the considered geometric configuration; (b) Diagrammatic representation of the bare Coulomb interations; (c) the RPA Dyson's equation for the screened external potential matrix.

Actually, for the early experiments observing partial field penetration, specially, in Ref. (2), the electric field is uniform, so T-F can be greatly simplified into analytical formula and adopted there. If non-uniformity is introduced, in spirit to the density functional theory (DFT) [8], T-F model has to be solved self-consistently [8, 9, 10]. Obviously it becomes not easy to extract the essential physics during the cumbersome numerical treatment. Furthermore a notable limitation of T-F model is that it is requires the density profile varies on length scales larger than the Fermi wavelength, i.e. $|\nabla n/n| < k_F$, where $k_F$ is the Fermi wave vector. It means, the celebrated Friedel Oscillation (FO) [11] cannot be accounted in T-F. Since FO is a major feature in long range, T-F is not very accurate if spatial non-uniformity is highlighted. An analytical treatment of this fundamental problem is still lacking. And we fill this gap in this paper.

*Theoretical Model.*- Our method is based on a linear screening model with multi-component random phase approximation (RPA). Concerning the bilayer system in Fig.1 (a), a negative point charge [12] with elemental charge $e_0$ is placed above the bilayer with a height $L$. The technique for the multicomponent system is used to derive the dynamically screened potential $U_{ij,scr}(q)$ in RPA [13]. The following equation is found, as illustrated diagrammatically in Fig.1(c):

$$U_{ij,scr}(q) = U_{ij,0}(q) + \sum_{l,m} V_{ijlm}(q)\Pi_{lm}(q)U_{lm,scr}(q) \quad (2)$$

where $q$: the wave vector parallel to layer plane; $V_{ijlm}(q) = \frac{e_0^2}{2\varepsilon_r q}\int dz \int dz' \xi_i^*(z)\xi_j(z)\xi_l^*(z')\xi_m(z')e^{-q|z-z'|}$: the bare coulomb interaction between electrons, which corresponds to the Feynman diagram in Fig.1(b); $U_{lm,0}(q)$: the bare/unscreened external potential; $\xi_i(z)$: the envelope wave function describing the confinement in $z$(vertical) direction for the $i$ layer; $i,j,l,m$ =1, 2 being the layer index, and we set upper(low) layer to be Layer 1(2) and these "1 or 2" subscript will be used throughout the paper; $\Pi$ is the irreducible non-interacting electronic 2D polarizability functions (i.e., the bare bubble in Fig. 1(c):

Eq.(2) is a self-consistent form and cannot give a direct formula of $U_{ij,scr}$. In this sense, it shows no improvement than T-F model. However, using the celebrated RPA dynamical dielectric function of multi-component system $\varepsilon_{ijlm}(q,\omega=0) = \varepsilon_{ijlm}(q) = \delta_{il}\delta_{jm} - V_{ijlm}(q)\Pi_{lm}(q)$ (Here frequency $\omega=0$ because we consider only the static screening, and $\omega$ will be omitted. $\delta_{il}$ is the usual Kronecker delta function) [13], we reduce it significantly to the convenient formula

$$U_{ij,scr}(q) = \sum_{l,m} \varepsilon^{-1}{}_{ijlm}(q)U_{lm,0}(q) \quad (3)$$

where $\varepsilon_{ijlm}(q)$ is the famous many-body formula defining the collective plasmon mode in an electron system.

We next address the issue of interlayer tunneling. As stated before, we are interested in those "spatially separated" screening, i.e. without inter-layer tunneling. This requires the inter-layer $V_{ijlm}$ term to be zero. From Fig.1 (b), $V_{ijlm}$ with $i=j$, $l=m$ stands for interlayer and intra-layer interaction. When $V_{ijlm}$ with $i \neq j$, $l \neq m$, like $V_{1211}$ or $V_{1212}$, it stands for the inter-layer tunneling term and we can set them to be zero. To fulfill this no interlayer tunneling, by neglecting the thickness of 2DEG, we assume the wave function $\xi_i(z)$ to be delta functions located at $z_i$ ($z_i$ being $i$ layer position in $z$ direction) [14]. Undoubtedly delta function can automatically meet the no interlayer tunneling requirement.

With $\xi_i(z)$ in hand, $U_{ij,0}(q)$ which being the Fourier transform of $U_{ij,0}(z) = \langle \xi_i(z)|U|\xi_j(z)\rangle$ can be calculated. Since this external potential $U$ is the coulomb potential exerted by a point charge, $U_{ij}(q)$ is easy to know by the two-dimensional Fourier transform of Coulomb interaction, leading to $U_{ij,0}(q) = \frac{e_0^2}{2\varepsilon_s q}\int \xi_i^*(z)\xi_j(z)\exp(-q|z_i - z_j|)dz$, so $U_{11,0}(q) = \frac{e_0^2}{2\varepsilon_s q}\exp(-qL)$, $U_{22,0}(q) = \frac{e_0^2}{2\varepsilon_s q}\exp[-q(L+d)]$, $U_{12,0}(q) = U_{21,0}(q) = 0$. Also from $V_{ijlm}(q)$ expression, we reach $V_{1111}(q) = V_{2222}(q) = \frac{e_0^2}{2\varepsilon_s q}$ and $V_{1122}(q) = V_{2211}(q) = \frac{e_0^2}{2\varepsilon_s q}\exp(-qd)$.

In this way, all left side terms in Eq.(3), including $\varepsilon_{ijlm}(q)$ and its inverse matrix $\varepsilon_{ijlm}^{-1}(q)$ is deduced straightforwardly. And we have: [15]

$$U_{ij,scr}(q) = \sum_{l,m} \varepsilon^{-1}{}_{ijlm}(q)U_{lm,0}(q) =$$

$$\begin{array}{c}11\\22\\12\\21\end{array}\begin{bmatrix} \frac{1-V_{2222}\Pi_{22}}{\varepsilon_{Intra}} & \frac{V_{1122}\Pi_{22}}{\varepsilon_{Intra}} & 0 & 0 \\ \frac{V_{1122}\Pi_{11}}{\varepsilon_{Intra}} & \frac{1-V_{1111}\Pi_{11}}{\varepsilon_{Intra}} & 0 & 0 \\ 0 & 0 & \frac{1-V_{1212}\Pi_{21}}{\varepsilon_{Inter}} & \frac{V_{1212}\Pi_{21}}{\varepsilon_{Inter}} \\ 0 & 0 & \frac{V_{1212}\Pi_{12}}{\varepsilon_{Inter}} & \frac{1-V_{1212}\Pi_{12}}{\varepsilon_{Inter}} \end{bmatrix}\begin{bmatrix} e^{-qL}\frac{e_0^2}{2\varepsilon_s q} \\ e^{-q(L+d)}\frac{e_0^2}{2\varepsilon_s q} \\ 0 \\ 0 \end{bmatrix} \quad (4.1)$$

$$\phantom{xxxx}11\phantom{xxxxx}22\phantom{xxxxx}12\phantom{xxxxx}21$$

with $\varepsilon_{Intra} = (1-V_{1111}\Pi_{11})(1-V_{2222}\Pi_{22}) - V_{1122}^2\Pi_{11}\Pi_{22}$ (4.2)

$\varepsilon_{Inter} = 1 - V_{1212}(\Pi_{12} + \Pi_{21})$ (4.3)

The array index are indicated in left and low side of Eq.(4.1). Screened potential in real space, i.e. $U_{ij,scr}(r)$, is deduced from $U_{ij,scr}(q)$ by: [16]

$$U_{ij,scr}(r) = (2\pi)^{-1}\int_0^\infty U_{ij,scr}(q)J_0(qr)qdq, \quad (5)$$

where $J_0(z)$ is the zero order Bessel function of the first kind. And the screened potential in upper (low) QW $U_{U,scr}$ ($U_{L,scr}$) is equal to $U_{11,scr}(r)$ ($U_{22,scr}(r)$).

*Back Screening.*- In following, using Eq. (4, 5), we calculate the screened potential in a GaAs/AlGaAs double quantum well (DQW) system, which consists two GaAs QW separated by an

AlGaAs barrier. An instructive screening configuration ( we call it as "back screening", i.e. the screener is in the back side of studied object), are considered, in which we set the upper QW electron density $n_1$ unchanged, and vary the inter-layer distance $d$ and the low QW electron density $n_2$. In this manner, the influences of backside layer on upper layer's screening properties are directly extracted. To guarantee the appearance of FO, the external charge is close to upper layer, i.e. $L$=5nm. And the polarizability of ordinary GaAs 2DEG is used: [17, 18]

$$\varepsilon(q) = 1 + \frac{s}{q}[1 - \theta(q - 2k_F)\sqrt{1 - (\frac{2k_F}{q})^2}] \quad (6)$$

with $s = 2/a_r$; $a_r$: semiconductor Bohr radius of GaAs radius; $k_F = \sqrt{2\pi n_s}$: the Fermi wave vector; $\theta$: Heaviside step function.

We first explain the large $d$ case. As expected, in Fig.2(a), for the screened upper QW potential $U_{U,scr}$, upper layer's FO (UFO) occurs in large radius, namely $r$>50nm. While for low QW, due to the large separation form the charge, the screened low QW potential $U_{L,scr}$ is very small and we have to plot it another figure [Fig. 2 (b)] for clear presentation. The $U_{L,scr}$ distribution becomes a large wave-pocket with spatial extension around 100nm, comparable with $d$. FO in low layer (LFO) is then not observable. Furthermore, seen in Fig.2 (a) (inset), $U_{L,scr}$ [let us concentrate on UFO and $U_{U,scr}(r=0)$] don't show any observable changes even the low QW electron density $n_2$ changes around 10 times, indicating that "back screening" scheme is not effective under large inter-layer separation. Due to this point, in Fig.2(c), a much small $d$=5nm is used to enhance coupling between QWs. Because of this small $d$, in addition to UFO, LFO also appears. More importantly, an increased $n_2$ reduces $|U_{U,scr}|$ in all radius $r$ range (especially for large $r$ where FO prevails, e.g. $r$>50nm),

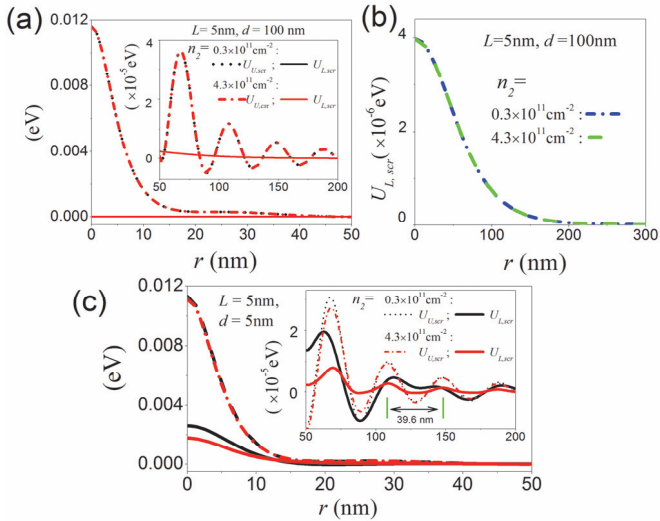

Fig.2 (color online). The spatial distribution of screened potential for upper ($U_{U,scr}$) and low ($U_{L,scr}$) QW in a GaAs/AlGaAs DQW system. And $n_1 = 1.0 \times 10^{11} \text{cm}^{-2}$. (a) and (b) $d$=100nm; (c) $d$=5nm. Since $U_{L,scr}$ in (a) is very small, they are also plot in (b) for clear presentation. The used $L$, $n_1$, $n_2$ values are indicated inset the figures.

hence demonstrating the validity of "back screening".

Another remarkable thing is the mutual influence between the period of UFO and LFO. We know that for a single GaAs 2DEG, $U_{scr}$ can be approximated by $\sim \sin(2k_F r)$, which is termed as FO and determines the FO period $\lambda = (\pi/k_F) \propto n^{-1}$. A naive thought is that the coupling may "increase" the electron density of upper layer, because low layer electrons can participate in the screening in upper layer. Then UFO period will become smaller if $n_2$ increases. We find this speculation is not very successful: UFO period is almost no changed with respect to increasing $n_2$. On the other hand, LFO period is subjected to a strong modulation from UFO. Shown in Fig.2 (c) inset, no matter what value $n_2$ is, an oscillation structure in $U_{L,scr}$ with period close to UFO period (∼39.6 nm) is present. The extreme case is $n_2 = 4.3 \times 10^{11} \text{cm}^{-2} > n_1$, in which $U_{L,scr}$ is totally dominated by UFO period, while LFO's own period (~19.1 nm) is not obviously observed.

*Effective polarizability Picture.-* Eq.(4) reveals an important rule that the screened potential $U_{scr}$ on the studied object, not only depends on its own properties, but also is influenced by the environment, like the polarizability of other objects nearby and their geometric placement. The central idea of this paper is to incorporate the environment's influences into the studied object. Therefore we only need to consider this modified "effective" object alone. This technique is mathematically summarized as:

$$U_{i,scr}(q) = \frac{U_{i,0}(q)}{\varepsilon_{i,eff}(q)} \quad (7)$$

where $\varepsilon_{i,eff}(q)$ is the effective dielectric function of object $i$, which contains the information of object $i$ plus the environment.

In case of our bilayer, for example, concerning upper layer, the unscreened potential there is $U_{U,0}(q) = U_{11}(q) = \frac{e_0^2}{2\varepsilon_s q}\exp(-qL)$.

Combining with Eq.(7), we reach:

$$\varepsilon_{U,eff}(q) = \frac{\varepsilon_{Intra}(q)}{1 - V_{2222}\Pi_{22}(1 - e^{-2qd})} \quad (8.1)$$

Then via the single-component system RPA expression $\varepsilon_{U,eff}(q) = 1 - V_{U,0}(q)\Pi_{U,eff}(q)$ with $V_{U,0}(q) = V_{1111}(q) = \frac{e_0^2}{2\varepsilon_s q}$, the effective polarizability $\Pi_U(q)$ is deduced to:

$$\Pi_{U,eff}(q) = \Pi_{11} + \frac{\Pi_{22}}{e^{2qd}(1 - V_{2222}\Pi_{22}) + V_{2222}\Pi_{22}} \quad (8.2)$$

With similar procedures, the effective dielectric function and polarizability of low layer is deduced to:

$$\varepsilon_{L,eff}(q) = \varepsilon_{Intra}(q) \quad (9.1)$$

$$\Pi_{L,eff}(q) = \Pi_{22} + \Pi_{11}[1 - V_{111}\Pi_{22}(1 - e^{-2qd})] \quad (9.2)$$

Then at $q$=0, it is approximated as:

$$\Pi_{U,eff}(q \to 0) = -[s_1 + \frac{s_2}{(1 + 2s_2 d)}]/(\frac{e_0^2}{2\varepsilon_s}) \quad (10.1)$$

$$\Pi_{L,eff}(q \to 0) \approx -[s_1 + s_2 + 2s_1 s_2 d]/(\frac{e_0^2}{2\varepsilon_s}) \quad (10.2)$$

We emphasize that Eq. (9.1) is exactly the dielectric function (although at frequency $\omega=0$) characterizing the collective plasmon excitation in two-component system [14]. From the condition required by plasmon occurrence, i.e. $\varepsilon_{Intra}(q)=0$, along with Eq. (9), we conclude that for multi-component systems, when plasmon occurs, electric field can not only reach any components, but also be amplified to infinite large in all reached components. We expect this viewpoint will also be true even $\omega \neq 0$. This field penetration viewpoint may help scientists catch the essences of plasmon excitation.

Compared with $\varepsilon_{eff}(q)$, $\Pi_{eff}(q)$ is more useful, because its cusp (or discontinuous points) directly reflects the FO period. Fig.3 displays the calculated $\Pi_{eff}(q)$ for the examples in Fig.(2). The features deserve attentions are the number of observable cusps. In Fig.3 (a) with $d$=5nm, for $n_2 = 0.3\times10^{11}\text{cm}^{-2} < n_1$, two cusps ($q=2k_{F1}$ and $q=2k_{F2}\approx 1.095 k_{F1}$) are observed in $\Pi_{U,eff}(q)$, which coincides with FO period in each layer. While for $n_2 = 4.3\times10^{11}\text{cm}^{-2} > n_1$, only one cusp is observed at $q=2k_{F1}$, because the $q=2k_{F2}\approx 4.147 k_{F1}$ cusp resides on a decreasing background of $\Pi_{U,0}(q)$ for $q>2k_{F1}$ region, where the electrons' rearranging (an alternative expression way of "screening") ability of upper QW is depressed. Therefore only upper layer's own cusp $q=2k_{F1}$ can survive in $\Pi_{U,eff}(q)$. Since for all $n_2$ values here, $q=2k_{F1}$ cusp is always the dominant cusp, it explains why UFO period is almost unchanged by varying $n_2$ in Fig.2(c). In Fig.3 (b), the above two cusps are both clearly observed in $\Pi_{L,eff}(q)$, which accounts for the complicated FO pattern in Fig.2(c), i.e. the simultaneously appearance of two FO with different period. On the question of which cusp is the sharpest or strongest, a simple rule can be summarized, i.e. its position is $q=2k_{Fm}$ with $k_{Fm}$ being the smaller one between $k_{F1}$ and $k_{F2}$. This determines which FO period is dominant in Fig.3(b) and explains the LFO results in Fig. 2(c) inset.

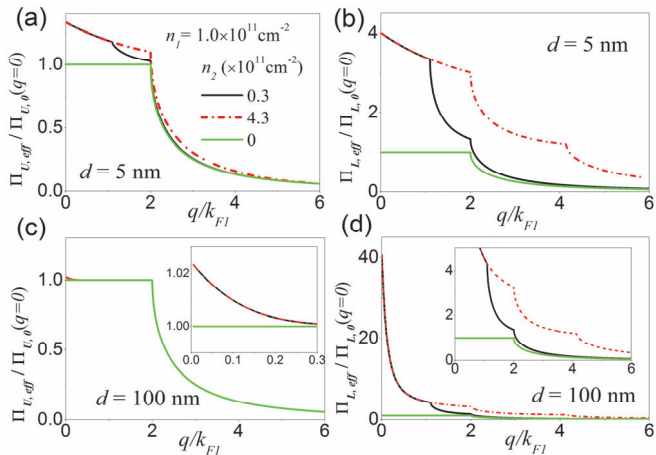

Fig. 3 (color online). The calculated effective polarizability for the example system mentioned in Fig.2. The related parameters are given inset. Note: $\Pi_{U,eff}(q)$ is normalized to $\Pi_{U,0}(q=0)$; by $k_F=\sqrt{2\pi n_s}$, $n_2 = 0.3, 4.3\times 10^{11}\text{cm}^{-2}$ corresponds $k_{F2}/k_{F1}\approx 0.548, 2.074$ respectively.

Next we turn to the large inter-QW separation $d$=100nm case. In Fig.3(c) for upper layer, we find $\Pi_{U,eff}(q)$ is almost not modified compared with $\Pi_{U,0}(q)$ except for small $q$ region. And only upper layer's own $q=2k_{F1}$ cusp appears. Interestingly, in Fig.3 (d) for low layer, $\Pi_{L,eff}(q)$ has been totally modified compared with $\Pi_{L,0}(q)$: much enhanced value and a different profile with the features(the cusps) being strongly smeared away.

Why in respect of screening, upper layer can influence low layer much stronger than the feed-back influence it receives? Via Fig.3(c, d), this enlightening phenomenon is quantitatively figured out by Eq.(9,10). A comparative illustration may help the readers capture this point. If upper layer is metal (i.e. in "front screening" configuration), no matter how close or far the low layer is, the screened potential in low layer is always zero. But in the reverse case (i.e. the "back screening" configuration), this rule is not true and the screening effect is limited by the inter-layer distance $d$, as discussed in below.

*Back screening with metal layer.*- After the preliminary discussion above, a heuristic problem naturally arises, which say: where is the limit of back screening? To address this issue, by varying backside layer electron density $n_2$, we calculate the evolution of screened potential in upper layer- $U_{U,scr}$. Two positions are specially chosen: $r$=100 nm or 0 nm which represent the position dominated by FO or not. $L$ is the same (i.e. 5nm). In Fig. 4(a), at $r$=0nm, with $n_2$ increases, $U_{U,scr}$ steadily decreases to approach a Limit I, which corresponding to an infinite large $n_2$. Limit I can be calculated by assuming $\varepsilon_{L,0}(q)=1+s_2/q$ for low GaAs QW. However, at $r$=100nm, FO makes difference and results a drastic oscillation pattern. So the naive belief that a larger $n_2$ always leads to an enhanced screening (i.e. smaller $U_{U,scr}$) is broken by FO at large radius. Even though, finally, limit I [although it is not the lowest $U_{U,scr}$ in Fig.4(b)] will be gradually reached with large enough $n_2$.

However Limit I is not the lowest [let us choose $U_{U,scr}(r=0)$ value as the criterion] screened potential we can reach. The ultimate screening limit is set by metal. The reason is simple: you cannot imagine one other material that has DOS larger than infinite large! To elucidate this point, we replace backside layer by a metal layer, forming a GaAs/AlGaAs 2DEG – metal bilayer system. We regard the metal layer as an ideal metal, which means $V_{2222}\Pi_{22}\to\infty$ [19, 20]. Most metal can well meet this requirement if frequency $\omega$ is below visible light frequency (of course, including the static case $\omega=0$ here). Therefore from Eq.(9,10), we reach :

$$\varepsilon_{U,eff}(q) = \varepsilon_{U,0}(q) + \frac{1}{(e^{2qd}-1)} \qquad (11.1)$$

$$\Pi_{U,eff}(q) = \Pi_{U,0}(q) + [-\frac{q}{(e^{2qd}-1)}]/(\frac{e_0^2}{2\varepsilon_s}) \qquad (11.2)$$

$$\Pi_{U,eff}(q\to 0) \approx -[s_1 + \frac{1}{2d}]/(\frac{e_0^2}{2\varepsilon_s}) \qquad (11.3)$$

Eq.(11) is strikingly important, indicating the distance $d$

between studied object and metal is the only factor on environment side that determines the screening. For instance, in the uniform case ($q=0$), metal can increase the studied object's DOS by a quantity proportional to $(2d)^{-1}$, which can be easily verified by T-F model [2]. Through Eq.(11), the calculated $U_{U,scr}$ (labeled as Limit II) is given in Fig. 4(a, b), and lower than Limit I as expected.

Metal screening is also characteristic by its ability to smoothen potential inhomogeneity. Shown in Fig.4(d), the backside metal can reduce the UFO amplitude by an order (~10%). The smoothening of UFO is closely related with the blunting of cusp in $\Pi_{U,eff}(q)$. From Eq.(11.2), the $\Pi_{U,eff}(q)$ part contributed by either metal or $n_2 = \infty$ case, is a continuous function without cusps. When mixing this part with studied object's original polarizability, the $q = 2k_{F1}$ cusp in resulted $\Pi_{eff}(q)$ becomes less sharp, as shown in Fig.4(c). Of course, metal layer is the strongest screener, thus most effectively damping $q = 2k_{F1}$ cusp.

Generally the smoothening ability in its nature is due to the diminished $U_{scr}$ via increased $\varepsilon_{U,eff}(q)$ [because of Eq.(7)]. Along this line, a natural argument is: an efficient smoothening of local potential fluctuation requires the dielectric function $\varepsilon_{U,eff}(q)$ contributed by the environment, e.g. 2D metal film here, to surpass the intrinsic dielectric function $\varepsilon_{U,0}(q)$ of the studied 2DEG. To support this argument, we consider the potential fluctuations with a typical size $\Delta r$, which corresponds to $q \sim 2\pi/\Delta r$. From Eq. (11.1), smoothening it needs the condition $(e^{2qd}-1)^{-1} > \varepsilon_{U,0}(q)$. If upper layer is an intrinsic single layer graphene (SLG) with average electron density being zero, thus $\varepsilon_{U,0}(q) = 1$ [4] and it immediately leads to $d <\sim \frac{\ln 2}{4\pi}\Delta r$. Neglecting the constant coefficient, we get $d < \Delta r$. It means for SLG, a metallic plate placed at distance $d$ can wipe out only electron-hole puddles (which is a special kind of potential fluctuation in graphene) with a typical size larger than $d$, as claimed in Ref.(6). However this allegation may only be correct for SLG. GaAs/AlGaAs 2DEG has some large polarizability, so in Fig.4 (d), though with FO size ~ 40nm>$d$=5nm, FO can still survive.

To conclude, we have shown that, to reflect the highly interacting nature, the screening in a coupled low-dimension systems can be viewed as the polarizability being modified by environment. Benefited by this highly heuristic and analytical evaluation method, many important fields can be re-examined handily, like the experimental studies on compressibility measurements of double Quantum Wires [21], screening of plasmons in graphene by semiconducting and metallic substrates [22], *etc*. In respect of experiments, we recently developed a heuristic method to introduce charges inside GaAs quantum well [23], which paves a way toward experimentally probing the potential inhomogeneity in charged low-dimension systems.

The work was supported by the Natural Science Foundation of China 11204334, the "Hundred Talent program" of the Chinese Academy of Sciences.

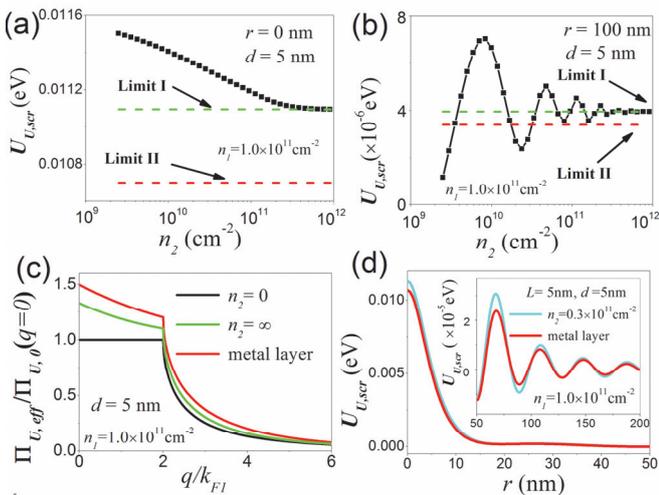

Fig. 4 (color online). Upper layer is a GaAs 2DEG with $n_1 = 1.0 \times 10^{11} cm^{-2}$: (a). Via an increasing $n_2$, the calaculated screened potential of upper layer $U_{U,scr}$ in a GaAs/AlGaAs DQW system. The position is at $r$=0nm; (b) Same as (a), but the position is at $r$=100nm; (c) effective polarizability for upper layer; (d) The spatial distribution of $U_{U,scr}$ when low layer is GaAs 2DEG or metal layer.